\documentclass[twocolumn,showpacs,aps,epsfig]{revtex4}

\usepackage{graphicx}
\usepackage{epstopdf}
\usepackage{latexsym}
\usepackage{amssymb}

\usepackage[center]{subfigure}

\begin{document}

 \newcommand{\bq}{\begin{equation}}
 \newcommand{\eq}{\end{equation}}
 \newcommand{\bqn}{\begin{eqnarray}}
 \newcommand{\eqn}{\end{eqnarray}}
 \newcommand{\nb}{\nonumber}
 \newcommand{\lb}{\label}
\newcommand{\PRL}{Phys. Rev. Lett.}
\newcommand{\PL}{Phys. Lett.}
\newcommand{\PR}{Phys. Rev.}
\newcommand{\CQG}{Class. Quantum Grav.}

\title{Black holes, compact objects and solar system tests in nonrelativistic  
general covariant theory of gravity}

\author{Jared Greenwald}
\email{Jared_Greenwald@baylor.edu}

\author{V. H. Satheeshkumar}
\email{VH_Satheeshkumar@baylor.edu}

\author{Anzhong Wang}
\email{Anzhong_Wang@baylor.edu}

\affiliation{GCAP-CASPER, Physics Department, Baylor
University, Waco, TX 76798-7316, USA }

\date{\today}

\begin{abstract}

We study spherically symmetric static spacetimes generally filled with an anisotropic fluid in the 
nonrelativistic general covariant theory of gravity. In particular, we find that the vacuum solutions 
are not unique, and can be expressed in terms of the $U(1)$ gauge field $A$. When solar system 
tests are considered, severe constraints on $A$ are obtained, which seemingly pick up the Schwarzschild 
solution uniquely.  In contrast to other versions of the Horava-Lifshitz theory, non-singular static 
stars made of a perfect fluid without heat flow can be constructed, due to the coupling of  the 
fluid with the gauge field.  These include the solutions with a constant pressure. We also study the 
general junction conditions across the surface of a star. In general, the conditions allow the existence 
of a thin matter shell on  the surface. When applying these conditions to the perfect fluid solutions 
with the vacuum ones as describing their external spacetimes, we find explicitly the matching 
conditions in terms of the parameters appearing in the solutions. Such matching is possible even 
without  the presence of a thin matter shell.

\end{abstract}


\maketitle

\section{Introduction}

\renewcommand{\theequation}{1.\arabic{equation}} \setcounter{equation}{0}

Recently, Horava proposed a  theory of quantum gravity \cite{Horava}, motivated by the Lifshitz 
theory in solid state physics \cite{Lifshitz}. 
Due to several remarkable features, the Horava-Lifshitz (HL) theory has  attracted a great deal 
of attention (see for example, \cite{HWW} and references therein).  

In the HL theory,  Lorentz symmetry is broken in the  ultraviolet (UV). The breaking manifests in the 
strong anisotropic scalings  between space and time, 
\bq
\lb{1.1}
{\bf x} \rightarrow \ell {\bf x}, \;\;\;  t \rightarrow \ell^{z} t.
\eq
In $(3+1)$-dimensional spacetimes,   the theory is power-counting 
renormalizable,   provided that  $z \ge 3$.   At low energies, the theory is expected to  flow 
to $z = 1$, whereby the Lorentz invariance is ``accidentally restored." 
Such an anisotropy between  time and space can be easily realized, when writing the metric
in the Arnowitt-Deser-Misner  (ADM) form  \cite{ADM},
 \bqn
 \lb{1.2}
ds^{2} &=& - N^{2}c^{2}dt^{2} + g_{ij}\left(dx^{i} + N^{i}dt\right)
     \left(dx^{j} + N^{j}dt\right), \nb\\
     & & ~~~~~~~~~~~~~~~~~~~~~~~~~~~~~~  (i, \; j = 1, 2, 3).~~~
 \eqn
 Under the rescaling (\ref{1.1})  
 with   $z  = 3$ (a condition we shall assume 
 in   this paper),   $N, \; N^{i}$ and $g_{ij}$ scale  as, 
 \bq
 \lb{1.3}
  N \rightarrow  N ,\;  N^{i}
\rightarrow {\ell}^{-2} N^{i},\; g_{ij} \rightarrow g_{ij}.
 \eq
The gauge symmetry of the system are the  foliation-preserving 
diffeomorphisms Diff($M, \; {\cal{F}}$),  
\bq
\lb{1.4}
\tilde{t} = t - f(t),\; \;\; \tilde{x}^{i}  =  {x}^{i}  - \zeta^{i}(t, {\bf x}),
\eq
for which $N, \; N^{i}$ and $g_{ij}$ transform as
\bqn
\lb{1.5}
\delta{g}_{ij} &=& \nabla_{i}\zeta_{j} + \nabla_{j}\zeta_{i} + f\dot{g}_{ij},\nb\\
\delta{N}_{i} &=& N_{k}\nabla_{i}\zeta^{k} + \zeta^{k}\nabla_{k}N_{i}  + g_{ik}\dot{\zeta}^{k}
+ \dot{N}_{i}f + N_{i}\dot{f}, \nb\\
\delta{N} &=& \zeta^{k}\nabla_{k}N + \dot{N}f + N\dot{f},
\eqn
where $\dot{f} \equiv df/dt$,  $\nabla_{i}$ denotes the covariant 
derivative with respect to the 3-metric $g_{ij}$, and  $N_{i} = g_{ik}N^{k}$, etc. From these expressions one can see that the
lapse function $N$ and the shift vector $N_{i}$ play the role of gauge fields of the Diff($M, \; {\cal{F}}$)
symmetry. Therefore, it is natural to assume that $N$ and $N_{i}$ inherit the same dependence on 
spacetime as the corresponding generators,
\bq
\lb{1.6}
N = N(t), \;\;\; N_{i} = N_{i}(t, x),  
\eq
which is   clearly preserved by the Diff($M, \; {\cal{F}}$), and usually referred to as the projectability condition. 

Due to the restricted diffeomorphisms (\ref{1.4}), one more degree of freedom appears
in the gravitational sector -  the spin-0 graviton. This is potentially dangerous, and needs to decouple  
in the infrared  (IR), in order to be consistent with observations.   Whether it is the case or not is still an open
question \cite{Mukreview}. 
In particular,  the spin-0 mode is not stable in the Minkowski background,
 in the original version of the HL theory  \cite{Horava} and in the Sotiriou, Visser 
and Weinfurtner (SVW) generalization \cite{SVW,WM}. Although in the SVW setup it is stable in the de Sitter
background   \cite{HWW}. In addition, non-perturbative analysis showed that 
it indeed decouples in the vacuum spherical static  \cite{Mukreview} and  cosmological \cite{WWb} spacetimes. 

To overcome the problem, various models have been proposed \cite{Sotiriou}. In particular, 
Horava and Melby-Thompson (HMT)  \cite{HMT} recently proposed a version in which  the spin-0 graviton is 
completely eliminated by introducing a Newtonian pre-potential $\varphi$ and a local U(1) gauge field
$A$, so that the foliation-preserving-diffeomorphisms, Diff($M, \; {\cal{F}}$), are extended to  
\bq
\lb{symmetry}
 U(1) \ltimes {\mbox{Diff}}(M, \; {\cal{F}}).
 \eq
Effectively, the spatial diffeomorphism symmetries of general relativity are kept intact, but  the
 time reparametrization symmetry   is contracted to a local gauge symmetry \cite{TH}. The restoration of general covariance, 
characterized by  Eq.(\ref{symmetry}),  nicely maintains the special status of time,  so that the anisotropic scaling (\ref{1.1})
with $z > 1$ can still be  realized.  Under the Diff($M, \; {\cal{F}}$),
the fields $A$ and $\varphi$ transform as,
\bqn
\lb{2.2}
\delta{A} &=& \zeta^{i}\partial_{i}A + \dot{f}A  + f\dot{A},\nb\\
\delta \varphi &=&  f \dot{\varphi} + \zeta^{i}\partial_{i}\varphi,
\eqn
while under the local $U(1)$,  they, together with $g_{ij}$,
 transform as
\bqn
\lb{2.3}
\delta_{\alpha}A &=&\dot{\alpha} - N^{i}\nabla_{i}\alpha,\;\;\;
\delta_{\alpha}\varphi = - \alpha,\nb\\ 
\delta_{\alpha}N_{i} &=& N\nabla_{i}\alpha,\;\;\;
\delta_{\alpha}g_{ij} = 0 = \delta_{\alpha}{N},
\eqn
where $\alpha$ is   the generator  of the local $U(1)$ gauge symmetry. 
A remarkable by-production of this ``non-relativistic general covariant" setup is that it forces the coupling constant
$\lambda$, introduced originally  to characterize the deviation of the kinetic part of the action from GR \cite{Horava},
to take exactly  its relativistic value $\lambda = 1$. A different view   can be found in \cite{Silva}.

In this paper,  we  investigate systematically the spherically symmetric spacetimes in the HMT
setup. In particular, after briefly   reviewing the theory in Sec. II, we develop the
general formulas of such spacetimes generally filled with an anisotropic fluid with heat flow in Sec. III. 
Then, in Sec. IV we study the vacuum solutions, and express them all in terms of the
gauge field $A$. Although the solutions are not unique, when we apply them to solar system 
tests in Sec. V, we find that these tests seemingly pick up the Schwarzschild solution generically. 
It should be noted that solar system tests were studied by several authors in the framework of the 
HL theory mainly without the projectability condition \cite{solar}. In Sec. VI,
we study  perfect fluid solutions without heat flow, and also express them all in terms of the gauge field
$A$.  By properly choosing $A$, non-singular stars can be constructed, due to the coupling of the gauge
field with the fluid. This is in contrast to all the previous versions of the
HL theory, in which it was shown that non-singular static  perfect fluid soltuons without heat
flow do not exist \cite{IM}, although the ones with heat fluid do \cite{GPW}. 
Then, we restrict ourselves to the cases where  the pressure is a constant, which is quite 
similar to the Schwarzshcild fluid solution. We show explicitly  that these solutions are 
free of spacetime singularities at the center of the star.
In Sec. VII we consider the junction conditions across the surface of a compact object with the minimal 
requirement that
the matching is mathematically meaningful. In particular, this allows the existence of a thin matter
shell on the surface of the star, where the pressures of the thin shell can have high-order derivatives of the
Dirac delta function. applying these conditions to the solutions found in Secs. V and VI, we obtain the
matching conditions with or without a thin matter shell. Finally, in Sec. VIII we present our main 
results and concluding remarks.  

It should be noted that spherically symetric static spacetimes in other versions of the HL theory have been
extensively studied, and it is difficult to give a complete list here. Instead, we simply refer readers
to references given in \cite{HWW}, and to the more recent review articles \cite{Mukreview,Sotiriou}.

\section{Nonrelativisitc general covariant HL theory}

\renewcommand{\theequation}{2.\arabic{equation}} \setcounter{equation}{0}

In this section, we shall  give a very brief introduction to the non-relativistic general covariant theory of gravity. For details, we refer
readers to  \cite{HMT}. We shall closely follow \cite{WW}, so that the notations and conversations will be used directly without
further explanations. In the following, \cite{WW} will be referred to as Paper I. 

The total action is given by,
 \bqn \lb{2.4}
S &=& \zeta^2\int dt d^{3}x N \sqrt{g} \Big({\cal{L}}_{K} -
{\cal{L}}_{{V}} +  {\cal{L}}_{{\varphi}} +  {\cal{L}}_{{A}} \nb\\
& & ~~~~~~~~~~~~~~~~~~~~~~ \left. +\frac{1}{\zeta^{2}} {\cal{L}}_{M} \right),
 \eqn
where $g={\rm det}\,g_{ij}$, and
 \bqn \lb{2.5}
{\cal{L}}_{K} &=& K_{ij}K^{ij} -   K^{2},\nb\\
{\cal{L}}_{\varphi} &=&\varphi {\cal{G}}^{ij} \Big(2K_{ij} + \nabla_{i}\nabla_{j}\varphi\Big),\nb\\
{\cal{L}}_{A} &=&\frac{A}{N}\Big(2\Lambda_{g} - R\Big).
 \eqn
Here   the coupling constant $\Lambda_{g}$, acting like a 3-dimensional cosmological
constant, has the dimension of (length)$^{-2}$. The 
Ricci and Riemann terms all refer to the three-metric $g_{ij}$.
 $K_{ij}$ is the extrinsic curvature, and ${\cal{G}}_{ij}$ is the 3-dimensional ``generalized"
Einstein tensor, defined  by
 \bqn \lb{2.6}
K_{ij} &=& \frac{1}{2N}\left(- \dot{g}_{ij} + \nabla_{i}N_{j} +
\nabla_{j}N_{i}\right),\nb\\
{\cal{G}}_{ij} &=& R_{ij} - \frac{1}{2}g_{ij}R + \Lambda_{g} g_{ij}.
 \eqn
${\cal{L}}_{M}$ is the
matter Lagrangian density and  
${\cal{L}}_{{V}}$  an arbitrary Diff($\Sigma$)-invariant local scalar functional
built out of the spatial metric, its Riemann tensor and spatial covariant derivatives, without the use of time derivatives. 
In \cite{SVW}, by assuming that the highest order derivatives are six and that  the theory  respects 
the parity, SVW constructed the most general form of  ${\cal{L}}_{{V}}$, given by
 \bqn \lb{2.5a} 
{\cal{L}}_{{V}} &=& \zeta^{2}g_{0}  + g_{1} R + \frac{1}{\zeta^{2}}
\left(g_{2}R^{2} +  g_{3}  R_{ij}R^{ij}\right)\nb\\
& & + \frac{1}{\zeta^{4}} \left(g_{4}R^{3} +  g_{5}  R\;
R_{ij}R^{ij}
+   g_{6}  R^{i}_{j} R^{j}_{k} R^{k}_{i} \right)\nb\\
& & + \frac{1}{\zeta^{4}} \left[g_{7}R\nabla^{2}R +  g_{8}
\left(\nabla_{i}R_{jk}\right)
\left(\nabla^{i}R^{jk}\right)\right],  ~~~~
 \eqn 
 where the coupling  constants $ g_{s}\, (s=0, 1, 2,\dots 8)$  are all dimensionless. The relativistic limit in the IR
 requires  \cite{SVW},
 \bq
 \lb{2.5b}
 g_{1} = -1,\;\;\; \zeta^2 = \frac{1}{16\pi G}.
 \eq
In Paper I,  this possibility was left open. To compare with the results obtained in \cite{GPW}, 
which will be referred to as Paper II, we shall 
restrict ourselves to these values.

Variation of the total action (\ref{2.4}) with respect to the lapse function $N(t)$  yields the
Hamiltonian constraint,
 \bq \lb{eq1}
\int{ d^{3}x\sqrt{g}\left({\cal{L}}_{K} + {\cal{L}}_{{V}} - \varphi {\cal{G}}^{ij}\nabla_{i}\nabla_{j}\varphi\right)}
= 8\pi G \int d^{3}x {\sqrt{g}\, J^{t}},
 \eq
where
 \bq \lb{eq1a}
J^{t} = 2 \frac{\delta\left(N{\cal{L}}_{M}\right)}{\delta N}.
 \eq
 
Variation of the action with respect to the shift $N^{i}$ yields the
super-momentum constraint,
 \bq \lb{eq2}
\nabla_{j}\Big(\pi^{ij} - \varphi  {\cal{G}}^{ij}\Big) = 8\pi G J^{i},
 \eq
where the super-momentum $\pi^{ij} $ and matter current $J^{i}$
are defined as
 \bqn \lb{eq2a}
\pi^{ij} &\equiv& \frac{\delta{\cal{L}}_{K}}{\delta\dot{g}_{ij}}
 = - K^{ij} +  K g^{ij},\nb\\
J^{i} &\equiv& - N\frac{\delta{\cal{L}}_{M}}{\delta N_{i}}.
 \eqn
Similarly, variations of the action with respect to $\varphi$ and $A$ yield, 
\bqn
\lb{eq4a}
& & {\cal{G}}^{ij} \Big(K_{ij} + \nabla_{i}\nabla_{j}\varphi\Big) = 8\pi G J_{\varphi},\\
\lb{eq4b}
& & R - 2\Lambda_{g} =    8\pi G J_{A},
\eqn
where
\bq
\lb{eq5}
J_{\varphi} \equiv - \frac{\delta{\cal{L}}_{M}}{\delta\varphi},\;\;\;
J_{A} \equiv 2 \frac{\delta\left(N{\cal{L}}_{M}\right)}{\delta{A}}.
\eq
On the other hand, variation with respect to $g_{ij}$ leads to the
dynamical equations,
 \bqn \lb{eq3}
&&
\frac{1}{N\sqrt{g}}\left[\sqrt{g}\left(\pi^{ij} - \varphi {\cal{G}}^{ij}\right)\right]_{,t} 
= -2\left(K^{2}\right)^{ij}+2K K^{ij}
\nb\\
& &  ~~~~~ + \frac{1}{N}\nabla_{k}\left[N^k \pi^{ij}-2\pi^{k(i}N^{j)}\right]\nb\\
& & ~~~~~
+  \frac{1}{2} \left({\cal{L}}_{K} + {\cal{L}}_{\varphi} + {\cal{L}}_{A}\right) g^{ij} \nb\\
& &  ~~~~~    + F^{ij} + F_{\varphi}^{ij} +  F_{A}^{ij} + 8\pi G \tau^{ij},
 \eqn
where $\left(K^{2}\right)^{ij} \equiv K^{il}K_{l}^{j},\; f_{(ij)}
\equiv \left(f_{ij} + f_{ji}\right)/2$, and
 \bqn
\lb{eq3a} 
F_{A}^{ij} &=& \frac{1}{N}\left[AR^{ij} - \Big(\nabla^{i}\nabla^{j} - g^{ij}\nabla^{2}\Big)A\right],\nb\\ 
F_{\varphi}^{ij} &=&  \sum^{3}_{n=1}{F_{(\varphi, n)}^{ij}},\nb\\
F^{ij} &\equiv& \frac{1}{\sqrt{g}}\frac{\delta\left(-\sqrt{g} {\cal{L}}_{V}\right)}{\delta{g}_{ij}}
 = \sum^{8}_{s=0}{g_{s} \zeta^{n_{s}} \left(F_{s}\right)^{ij} },\nb\\
 \eqn
with 
$n_{s} =(2, 0, -2, -2, -4, -4, -4, -4,-4)$.  The geometric 3-tensors $ \left(F_{s}\right)_{ij}$ and 
$F_{(\varphi, n)}^{ij}$ are given by Eqs.(2.21)-(2.23)
in Paper I, which, for the sake of the readers' convenience, are reproduced in  Eqs.(\ref{A.1}) and (\ref{A.2}) of this
paper. The stress 3-tensor $\tau^{ij}$ is defined as
 \bq \label{tau}
\tau^{ij} = {2\over \sqrt{g}}{\delta \left(\sqrt{g}
 {\cal{L}}_{M}\right)\over \delta{g}_{ij}}.
 \eq
 
The matter quantities $(J^{t}, \; J^{i},\; J_{\varphi},\; J_{A},\; \tau^{ij})$ satisfy the
conservation laws,
 \bqn \lb{eq5a} & &
 \int d^{3}x \sqrt{g} { \left[ \dot{g}_{kl}\tau^{kl} -
 \frac{1}{\sqrt{g}}\left(\sqrt{g}J^{t}\right)_{, t}  
 +   \frac{2N_{k}}  {N\sqrt{g}}\left(\sqrt{g}J^{k}\right)_{,t}
  \right.  }   \nb\\
 & &  ~~~~~~~~~~~~~~ \left.   - 2\dot{\varphi}J_{\varphi} -  \frac{A} {N\sqrt{g}}\left(\sqrt{g}J_{A}\right)_{,t}
 \right] = 0,\\
\lb{eq5b} & & \nabla^{k}\tau_{ik} -
\frac{1}{N\sqrt{g}}\left(\sqrt{g}J_{i}\right)_{,t}  - \frac{J^{k}}{N}\left(\nabla_{k}N_{i}
- \nabla_{i}N_{k}\right)   \nb\\
& & \;\;\;\;\;\;\;\;\;\;\;- \frac{N_{i}}{N}\nabla_{k}J^{k} + J_{\varphi} \nabla_{i}\varphi - \frac{J_{A}}{2N} \nabla_{i}A
 = 0.
\eqn

\section{ Spherically Symmetric Static Spacetimes}

\renewcommand{\theequation}{3.\arabic{equation}} \setcounter{equation}{0}

Spherically symmetric static spacetimes in the framework of the SVW setup are studied systematically in Paper II.
In this section, we shall closely follow the development presented there. 
In particular, the metric for  static spherically symmetric spacetimes that preserve the form of Eq. (\ref{1.2}) 
with the projectability condition can be cast in the form,  
\bq
\lb{3.1b}
ds^{2} = - c^{2}dt^{2} + e^{2\nu} \left(dr + e^{\mu - \nu} dt\right)^{2}  + r^{2}d^2\Omega,  
\eq
in the spherical coordinates $x^{i} = (r, \theta, \phi)$, where  $d^2\Omega = d\theta^{2}  + \sin^{2}\theta d\phi^{2}$,
and 
\bq
\lb{3.1}
 \mu = \mu(r),\;\;\; \nu = \nu(r),\;\;\; N^{i} = \left\{e^{\mu - \nu}, 0, 0\right\}.
 \eq
The corresponding timelike Killing vector is  $\xi = \partial_{t}$. For the  above metric,  one finds
\bqn
\lb{3.3a}
K_{ij} &=& e^{\mu+\nu}\Big(\mu'\delta^{r}_{i}\delta^{r}_{j} + re^{-2\nu}\Omega_{ij}\Big),\nb\\
R_{ij} &=&  \frac{2\nu'}{r}\delta^{r}_{i}\delta^{r}_{j} + e^{-2\nu}\Big[r\nu' - \big(1-e^{2\nu}\big)\Big]\Omega_{ij},\nb\\
{\cal{L}}_{K} &=& - \frac{2}{r^{2}} e^{2(\mu-\nu)}\left(2r\mu' + 1\right),  \nb\\
{\cal{L}}_{\varphi} &=& \frac{\varphi e^{-4\nu}}{r^2} \Bigg\{\Big[e^{2\nu}\left(\Lambda_{g} r^2 - 1\right) + 1 \Big]
    \Big(\varphi '' - \nu'\varphi' \nb\\
    & & + 2e^{\mu + \nu}\mu'\Big) - 2\Big(\nu' -  \Lambda_{g} re^{2\nu}\Big)\Big(\varphi' + 2e^{\mu +\nu}\Big)\Bigg\},   \nb\\
{\cal{L}}_{A} &=&  \frac{2A}{r^2} \Big[e^{-2\nu}\left(1 - 2r \nu'\right) + \left(\Lambda_{g} r^2 - 1\right)\Big],\nb\\
{\cal{L}}_{V} &=& \sum_{s=0}^{3}{{\cal{L}}_{V}^{(s)}},
\eqn
where  a prime denotes the ordinary derivative with respect to its indicated argument,
  $\Omega_{ij} \equiv \delta^{\theta}_{i}\delta^{\theta}_{j}  + \sin^{2}\theta\delta^{\phi}_{i}\delta^{\phi}_{j}$,
  and ${\cal{L}}_{V}^{(s)}$'s are given by Eq.(A1) in Paper II. 
 Then,  the Hamiltonian constraint (\ref{eq1}) reads,
 \bq 
 \lb{3.3b}
\int{\left( {\cal{L}}_{K} + {\cal{L}}_{V} - {\cal{L}}_{\varphi}^{(1)} - 8 \pi G J^{t} \right) e^{\nu} r^{2} dr}
= 0,
 \eq 
 where 
\bqn
{\cal{L}}_{\varphi}^{(1)} &=& \frac{\varphi e^{-4\nu}}{r^2} \Bigg\{\Big[e^{2\nu}\left(\Lambda_{g}r^2 -1\right) + 1\Big] (\varphi'' - \nu' \varphi')\nb\\
& &~~~~~~~~~~~  - 2\left(\nu' -    \Lambda_{g} re^{2\nu}\right) \varphi'\Bigg\},  
\eqn 
 while the momentum constraint (\ref{eq2}) yields,
 \bqn
 \lb{3.3c}
  & & 2 r \nu' + e^{-(\mu + \nu)} \Big[e^{2\nu}\left(\Lambda_{g}r^2 - 1\right) + 1 \Big] \varphi' \nb\\ 
  & & ~~~~~~~~~~~~~~~~~~~~~  =  - 8\pi G r^2 e^{2(\nu -\mu)}  v,
 \eqn
 where 
 $J^{i} = e^{-(\mu + \nu)}\big(v, 0, 0\big)$.
 It can be also shown that Eqs.(\ref{eq4a}) and (\ref{eq4b}) now read,
 \bqn
 \lb{3.3e}
& &  \Big[e^{2\nu}\left(\Lambda_{g}r^2 -1\right) + 1\Big]\Big(\varphi'' - \nu' \varphi' + e^{\mu + \nu}\mu'\Big)\nb\\
&& ~ -2\Big(\nu' - \Lambda_{g}re^{2\nu}\Big)\Big(\varphi' + e^{\mu + \nu}\Big)
= 8\pi G r^{2} e^{4\nu} J_{\varphi}, ~~~~~~~~ \\
\lb{3.3f}
& & 2 r \nu'  - \Big[e^{2\nu}\left(\Lambda_{g}r^2 - 1\right) +  1\Big] =   4\pi G r^2 e^{2\nu}  J_{A}.
\eqn
 The dynamical equations (\ref{eq3}), on the other hand, yield,
 \bqn
 \lb{3.3g}
& &2\big(\mu' + \nu'\big) + \frac{1}{r} + \frac{1}{2}re^{2(\nu-\mu)}\left({\cal{L}}_{\varphi} + {\cal{L}}_{A}\right) \nb\\
 & &  ~ = - re^{-2\mu}\Big(F_{rr} + F^{\varphi}_{rr} + F^{A}_{rr} + 8\pi G e^{2\nu}p_{r}\Big),~~~~\\
 \lb{3.3h}
 & &  \mu'' + \big(2\mu' - \nu'\big)\left(\mu' + \frac{1}{r}\right)  + \frac{1}{2}e^{2(\nu-\mu)}\left({\cal{L}}_{\varphi} + {\cal{L}}_{A}\right) \nb\\
 & &  ~  = -\frac{e^{ 2(\nu -\mu)}}{r^{2}}\Bigg(F_{\theta\theta} + F^{\varphi}_{\theta\theta} + F^{A}_{\theta\theta} + 8\pi G r^{2}p_{\theta}\Bigg), ~~~~
 \eqn
 where
 \bqn
 \lb{3.3i}
 \tau_{ij} &=& e^{2\nu}p_{r}\delta^{r}_{i}\delta^{r}_{j} + r^{2}p_{\theta}\Omega_{ij},\nb\\ 
F^{A}_{ij} &=&\frac{ 2}{r}\big(A' + A\nu'\big) \delta^{r}_{i}\delta^{r}_{j}  +  e^{-2\nu}\Big[r^{2}\big(A'' - \nu'A'\big)\nb\\
& & ~~~ + r\big(A' + A\nu'\big)  - A\Big(1 - e^{2\nu}\Big)\Big]\Omega_{ij},
\eqn
 $F_{ij}$'s for the metric  (\ref{3.1b}) are given by Eq.(A2) in Paper II,  and  $F_{(\varphi,s)}^{ij}$ are given by Eq.(\ref{B.2}) in
  the present paper.  
As in Paper II, here we define a fluid with $p_{r} = p_{\theta}$
as a perfect fluid, which in general conducts heat flow along the radial direction \cite{Santos85}. 
 
Since the spacetime is static, one can see that now the energy conservation law (\ref{eq5a}) is satisfied identically,
while the momentum conservation (\ref{eq5b}) yields,
\bq
\lb{3.3j}
v\mu' - \big(v' - p_{r}'\big) - \frac{2}{r}\big(v - p_{r} + p_{\theta}\big) + J_{\varphi} \varphi' - \frac{1}{2}J_{A} A'   = 0.
\eq

To relate   the quantities $J^{t},\; J^{i}$ and $\tau_{ij}$   to the ones often used in general relativity,  following 
Paper II, one can first introduce the unit normal vector $n_{\mu}$ to the hypersurfaces $t =$ Constant, and then 
the spacelike unit vectors $\chi_{\mu}, \; \theta_{\mu}$ and
$\phi_{\mu}$, defined as
 \cite{Ann}
\bqn
\lb{emt1}
 n_{\mu} &=& \delta^{t}_{\mu}, \;\;\; n^{\mu} = - \delta_{t}^{\mu} + e^{\mu-\nu}\delta_{r}^{\mu},\nb\\
 \chi^{\mu} &=& e^{-\nu}\delta^{\mu}_{r} , \;\;\; \chi_{\mu} = e^{\mu}\delta^{t}_{\mu} + e^{\nu} \delta^{r}_{\mu},\nb\\
 \theta_{\mu} &=& r\delta^{\theta}_{\mu},\;\;\; \phi_{\mu} = r\sin\theta \delta^{\phi}_{\mu}.
\eqn
In terms of these four unit vectors,    
the energy-momentum tensor for an anisotropic fluid with heat flow can be written as
\bqn
\lb{emt2}
T_{\mu\nu} &=& \rho_{H}n_{\mu}n_{\nu} + q \big(n_{\mu} \chi_{\nu} + n_{\nu} \chi_{\mu} \big)\nb\\
& & + p_{r}\chi_{\mu} \chi_{\nu}  + p_{\theta}\big(\theta_{\mu}\theta_{\nu} + \phi_{\mu}\phi_{\nu}\big),
\eqn
where $\rho_{H}, \; q,\; p_{r}$ and $p_{\theta}$ denote, respectively, the energy density, heat flow 
along radial direction, radial, and tangential pressures, measured by the observer with the four-velocity
$n_{\mu}$. 
Then,   one can see that such a decomposition is consistent with 
 the quantities $J^{t}$ and $J^{i}$, defined by
 \bq
 \rho_{H} = -2 J^{t},\;\;\; v= e^{\mu} q. 
\eq
  It should be noted that the definitions of the energy density $\rho_H$, 
 the radial pressure $p_r$ and the heat flow $q$ are different from the ones defined in a comoving frame 
 in general relativity. For detail,  we refer readers to Appendix B of Paper II.

 Finally, we note that in writing the above equations, we leave the choice of the $U(1)$ gauge open. From
 Eq.(\ref{2.3}) one can see that it can be used to set  one (and only one) of the three functions $A,\; \varphi$ 
 and $N_{r}$  to zero.  To compare our results  with the one obtained in \cite{HMT}, in the
 rest of this paper (except the first part of Sec. VII),  without loss of the generality, we shall choose the gauge,
 \bq
 \lb{gauge}
 \varphi =0.
 \eq
Then, we find that
\bq
\lb{Lphi}
{\cal{L}}_{\varphi} = 0, \;\;\; F^{ij}_{(\varphi, n)} = 0, \; (n = 1, 2, 3).
\eq

\section{Vacuum Solutions}

\renewcommand{\theequation}{4.\arabic{equation}} \setcounter{equation}{0}
 
 In the vacuum case, we have $J^t = v = p_{r} = p_{\theta} = J_{A} = J_{\varphi} = 0$.
 With the gauge (\ref{gauge}), from the momentum constraint (\ref{3.3c})   we immediately obtain
 $\nu = $ Constant, 
 while Eq. (\ref{3.3f}) further requires
 \bq
 \lb{4.1}
 \nu = 0,\;\;\; \Lambda_{g} = 0.
 \eq
This is different from  the solutions presented  in \cite{HMT}, where $\nu \not= 0,\; \mu = -\infty$.
Inserting the above into Eq.(\ref{3.3e}), it can be shown that it is satisfied identically. Since $\nu = 0$, from the expressions
of $\left(F_{s}\right)_{ij}$ given by Eq.(A2) in Paper II, we find that  $\left(F_{s}\right)_{ij} = 0$ for $s \not=0$,  and
$\left(F_{0}\right)_{ij} = -g_{ij}/2$, so that 
\bq
\lb{4.2}
F_{ij} = - \Lambda g_{ij},
\eq
where $\Lambda \equiv \zeta^2 g_0/2$.
Substituting Eqs.(\ref{3.3i}) and (\ref{Lphi}) - (\ref{4.2}) into Eqs.(\ref{3.3g}) and (\ref{3.3h}), we find that
\bqn
\lb{4.3a}
& & \Big(2r\mu' + 1\Big) e^{2\mu} = \Lambda r^{2} - 2r A',\\
\lb{4.3b}
& & \mu'' + 2\mu'\Big(\mu' + \frac{1}{r}\Big) =  \frac{e^{-2\mu}}{r}\Big[\Lambda r  - \big(rA'\big)'\Big].
\eqn
It can be shown that Eq.(\ref{4.3b}) is not independent, and can be obtained from Eq.(\ref{4.3a}). Therefore,
the solutions are not uniquely determined, since now we have only one equation, (\ref{4.3a}), for  two unknowns, 
$\mu$ and $A$. In particular, for any given $A$, from Eq.(\ref{4.3a}) we find that
\bq
\lb{4.4}
\mu = \frac{1}{2}\ln\Bigg[\frac{2m}{r} + \frac{1}{3}\Lambda r^2 - 2A(r) + \frac{2}{r}\int^{r}{A(r') dr'}\Bigg].
\eq
On the other hand, we also have 
\bq
\lb{4.4a}
{\cal{L}}_{K} = 
 \frac{4A'}{r} - 2\Lambda,\;\;\;
{\cal{L}}_{V} = 2 \Lambda.
\eq
Inserting it into the Hamiltonian constraint (\ref{3.3b}), we find that
\bq
\lb{4.4c}
\int_{0}^{\infty}{r A'(r) dr} = 0.
\eq
Therefore, for any given function $A$, subjected to the above constraint, the solutions given by Eqs.(\ref{4.1}) 
and (\ref{4.4}) represent the vacuum solutions of the HL theory.  Thus, in contrast to general relativity, the
vacuum solutions in the HMT setup are not unique.

When $A$ is a constant (without loss of generality, we can set $A = 0$), from the above we find that
\bq
\lb{4.4b}
 \mu = \frac{1}{2}\ln\left(\frac{2m}{r} + \frac{1}{3}\Lambda r^2 \right),\; (A = 0),
\eq
which is exactly the Schwarzschild (anti-) de Sitter solution, written in the Gullstrand-Painleve coordinates \cite{GP}.
It is interesting to note that when $m = 0$  we must assume that $\Lambda > 0$, in order to have $\mu$ real. 
That is, the anti-de 
Sitter solution cannot be written in the static ADM form (\ref{3.1b}).

\section{Solar System Tests}

\renewcommand{\theequation}{5.\arabic{equation}} \setcounter{equation}{0}
 
 The solar system tests are usually written in terms of the Eddington parameters, by following the so-called
 ``parameterized post-Newtonian" (PPN) approach, introduced initially by Eddington \cite{Edd}. The gravitational
 field, produced by a  point-like and motion-less particle with mass $M$,  is often described by the  
 form of metric \cite{Hartle},
 \bq
 \lb{6.1}
 ds^{2} = - e^{2\Psi}c^{2}dt^2 + e^{2\Phi}dr^2 + r^{2}d^2\Omega,
 \eq
 where $\Psi$ and $\Phi$ are functions of the dimensionless quantity $\chi \equiv GM/(rc^2)$ only. For the solar
 system, we have $GM_{\bigodot}/c^2 \simeq 1.5km$, so that in most cases we have $\chi \ll 1$. Expanding
 $\Psi$ and $\Phi$ in terms of  $\chi$, we have \cite{Hartle}
 \bqn
 \lb{6.2}
 e^{2\Psi} &=& 1 - 2\left(\frac{GM}{c^2 r}\right) + 2\big(\beta - \gamma\big) \left(\frac{GM}{c^2 r}\right)^2 + ...,\nb\\
 e^{2\Phi} &=& 1 + 2\gamma \left(\frac{GM}{c^2 r}\right) +  ..., 
 \eqn
where $\beta$ and $\gamma$ are the Eddington parameters. General relativity predicts $\beta = 1 = \gamma$
strictly, while the current radar ranging of the Cassini probe \cite{BT}, and the procession of lunar laser ranging data
\cite{WTB} yield, respectively, the bounds \cite{RJ},
\bqn
\lb{6.3}
\gamma - 1&=& (2.1 \pm 2.3)\times 10^{-5},\nb\\
\beta - 1&=& (1.2 \pm 1.1)\times 10^{-4},
\eqn
which are consistent with the predictions of general relativity. 

To apply the solar system tests to the HL theory, we need first to transfer the above bounds to the metric coefficients $\mu$
and $\nu$. In Appendix B of Paper II, the relations between $(\Phi,\; \Psi)$ and $(\mu, \; \nu)$ have been worked out explicitly,
and are given by
\bq
\lb{6.4}
\mu = \frac{1}{2}\ln\Bigg[c^{2}\Big(1 - e^{2\Psi}\Big)\Bigg],\;\;\; \nu = \Phi + \Psi,
\eq
or inversely,
\bqn
\lb{6.5}
\Phi &=& \nu - \frac{1}{2}\ln\Big( 1 - \frac{1}{c^{2}}e^{2\mu}\Big),\nb\\
\Psi &=&   \frac{1}{2}\ln\Big( 1 - \frac{1}{c^{2}}e^{2\mu}\Big).
\eqn 
Inserting Eq.(\ref{6.2}) into Eq.(\ref{6.4}), we find that
\bqn
\lb{6.6}
\mu &=& \frac{1}{2}\ln\left\{2c^{2}\Bigg[\left(\frac{GM}{c^{2}r}\right) - \big(\beta - \gamma\big)\left(\frac{GM}{c^{2}r}\right)^{2} + ...\Bigg]\right\},\nb\\
\nu &=& \big(\gamma - 1\big)  \left(\frac{GM}{c^{2}r}\right)  + ... .
\eqn
Comparing Eq.(\ref{6.6}) with Eq.(\ref{4.4}) for $\Lambda = 0$, we find that in order to be consistent with solar system tests, we must assume that
\bq
\lb{6.7}
A(r) = {\cal{O}}\left[\left(\frac{GM}{c^{2}r}\right)^2\right].
\eq
Together with the Hamiltonian constraint (\ref{4.4c}), we find that this is impossible unless 
$A = 0$.
 Therefore, although the vacuum solution in the HMT setup is not unique, the solar system tests seemingly require that it must be the Schwarzschild
vacuum solution. 

It should be noted that by choosing $A(r)$ in very  particular forms, the  condition $A = 0$ could be relaxed \cite{AP}. But, such chosen $A$ is not analytic (in terms
of the dimensionless quantity $\chi$), and it is not clear how to  expand it in the form of (\ref{6.6}). Thus,   in this paper we simply discard those possibilities.

\section{Perfect Fluid Solutions}

\renewcommand{\theequation}{6.\arabic{equation}} \setcounter{equation}{0}

In this section, let us consider   perfect fluid without heat flow, that is,
\bq
\lb{5.1}
p_{r} = p_{\theta} = p,\;\;\; v = 0.
\eq
Then, together with the gauge choice (\ref{gauge}), from Eq.(\ref{3.3c}) we find that $\nu =$ Constant. 
 However,
to be matched with the vacuum solutions outside of the fluid, as shown in Sec. IV, we must set this constant to  zero,
\bq
\lb{5.2}
\nu = 0,
\eq
from which we immediately find that $R_{ij} = 0$,  $F_{ij}$ is still given by Eq.(\ref{4.2}), and
\bqn
\lb{5.3}
{\cal{L}}_{A} &=& 2\Lambda_{g}A,\nb\\
F_{ij}^{A} &=& \frac{ 2A'}{r} \delta^{r}_{i}\delta^{r}_{j}  + r\big(rA'\big)'  \Omega_{ij}.
\eqn
Inserting the above into Eqs.(\ref{3.3e})-(\ref{3.3h}), we find that
\bqn
\lb{5.3aa}
& & J_{\varphi} = \frac{\Lambda_{g}}{8\pi G r^{2}}\left(r^{2}e^{\mu}\right)_{,r},\\
\lb{5.3ab}
& & J_{A} = -\frac{\Lambda_{g}}{4\pi G},\\
\lb{5.3a}
& & \big(rf\big)' + 2rA' + \Lambda_{g}r^2 A - \Lambda r^2 = -8\pi G r^2 p, ~~\\
\lb{5.3b}
& & \frac{1}{2}r f'' + f'  + \big(rA'\big)' + \Lambda_{g}r A - \Lambda r = -8\pi G r p,   ~~ 
\eqn
where $f \equiv e^{2\mu}$. From the last two equations, we find that
\bq
\lb{5.3c}
r^2 f'' -2 f  = -2r^3 \left(\frac{A'}{r}\right)'.
\eq
 On the other hand, the conservation law of momentum (\ref{3.3j}) now reduces to
 \bq
 \lb{5.4}
 p' + \frac{\Lambda_{g}}{8\pi G} A' = 0,
 \eq
 which has the solution,
 \bq
 \lb{5.5}
 p =   p_{0} - \frac{\Lambda_{g}}{8\pi G} A,
 \eq
where $p_{0}$ is an integration constant. 

Substituting it into Eq.(\ref{5.3a}), and then taking a 
derivative of it, we find that the resulting equation is exactly given by Eq.(\ref{5.3c}). 
Thus, both Eqs.(\ref{5.3c}) and (\ref{5.3b}) are not independent, and can all be derived from Eqs.(\ref{5.3a})
and (\ref{5.5}). Then, in the present case there are   five independent equations, the Hamiltonian 
constraint (\ref{3.3b}), and Eqs.(\ref{5.3aa}), (\ref{5.3ab}), (\ref{5.3a}) and (\ref{5.5}). 
However,  we have six unknowns, $A,\; \mu,\; p,\; J^{t},\; J_{\varphi},\; 
J_{A}$. Therefore, the problem now is not uniquely determined. As in the vacuum case, we
can express all these quantities in terms of the gauge field $A$. In particular, substituting Eq.(\ref{5.5})
into Eq.(\ref{5.3a}) and then integrating it, we obtain 
 \bq
 \lb{5.7a}
\mu = \frac{1}{2}\ln\Bigg[\frac{2m}{r} + \frac{1}{3}\Big(\Lambda - {8\pi G p_{0}}\Big)r^2 - 2A 
+ \frac{2}{r}\int^{r}{A(r') dr'}\Bigg].
 \eq
Then, we find that
\bq
\lb{5.6a}
{\cal{L}}_{\varphi} = \frac{4A'}{r} - 2\left(\Lambda - 8\pi G p_{0}\right),\;\;\;
{\cal{L}}_{V} = 2\Lambda.
\eq
Inserting the above into Eq.(\ref{3.3b}), we find that it can be cast in the form,
\bq
\lb{5.7c}
\int^{\infty}_{0}{\tilde{\rho}(r) dr} = 0,
\eq
where
\bq
\lb{5.7d}
J^{t} = \frac{1}{2\pi G}\Bigg(4\pi G p_{0} + \frac{A'(r)}{r} - \frac{\tilde{\rho}(r)}{r^2}\Bigg).
\eq

From the above one can see that once $A$ is given, one can immediately obtain all the rest. By properly
choosing it (and $\tilde{\rho}(r)$), it is not difficult to see that one can construct non-singular
solutions representing stars made of a perfect fluid without heat flow. To see this explicitly, 
let us consider the following two particular cases. 
 
 \subsection{$\Lambda_{g} = 0$}
 
 When $\Lambda_{g} = 0$, we have
\bq
\lb{5.7g}
J_{\varphi} = J_{A} = 0,\;\;\; p = p_{0},
\eq
while $\mu$ and $J^{t}$ are still given by Eqs.(\ref{5.7a}) and (\ref{5.7d}), respectively.
To have a physically acceptable model, we require that the fluid be non-singular in the center. Since $R_{ij} = 0$, one can see that any
quantity built from the Riemann and Ricci tensor vanishes in the present case. Then, possible singularities can only come from the kinetic part,
  $K_{ij}$, where the very first quantity   is 
\bqn
\lb{5.7e}
K &=& g^{ij}K_{ij} = \frac{e^{\mu}}{r}\left(r\mu' +2\right)\nb\\
&=& \frac{e^{-\mu}}{r}\left(\frac{3m}{r} + \left(\Lambda - 8\pi Gp_{0}\right)r^{2} - rA' - 3A\right.\nb\\
& & ~~~~~~~~~~~~~~~ \left. +\frac{3}{r}\int{A(r')dr'}\right).
 \eqn
Assuming that near the center $A$ is dominated by the term $r^{\alpha}$, we find that $K$ is non-singular only when
\bq
\lb{5.7f}
m = 0,\;\;\; \alpha \ge 2.
\eq
For such a function $A$, Eq.(\ref{5.7d}) show that $J^{t}$ is non-simgular, as long as $\tilde{\rho}(r) \simeq {\cal{O}}(r^{2})$.

 \subsection{$A = A_{0}$}
 
 When $A$ is a constant,    from Eq.(\ref{5.5}) we can see that the pressure $p$ is also a constant. Then, the integration of Eq.(\ref{5.3a}) yields,
  \bq
 \lb{5.7b}
\mu = \frac{1}{2}\ln\Bigg\{\frac{2m}{r} + \frac{1}{3}\Big(\Lambda   - {8\pi G p_{0}}\Big)r^2\Bigg\}.
 \eq
Inserting it into Eq.(\ref{5.3c}) we find that it is satisfied identically, while the Hamiltonian constraint (\ref{3.3b}) can also be cast in the form
of Eq.(\ref{5.7c}), but now with  
\bq
\lb{5.7e}
J^{t} = \frac{1}{8\pi G}\Bigg(16\pi G p_{0}    - \frac{\tilde{\rho}(r)}{r^2}\Bigg).
\eq
Thus, the solutions of Eqs.(\ref{5.2}) and (\ref{5.7b}) represent 
a perfect fluid with a constant pressure, $p = p(A_{0})$. 
In this case, it can be shown that $K$ is free of any spacetime singularity at the center only when $m = 0$. 

It should be noted that in \cite{IM} it was shown that
non-singular static solutions of perfect fluid without heat flow do not exist. Since their conclusions
only come from the conservation law of momentum, one might expect that this is also true in the 
current setup. However,  from Eq.(\ref{3.3j}) we can see that in the present case the conservation law
contains  two extra terms, $J_{\varphi}$ and $J_{A}$.  Only when both of them vanish, can one obtain the above
conclusions. Since in general one can only choose one of them to be zero by using the  gauge freedom,
it is expected that non-singular static stars can be constrcuted by properly choosing the gauge field $A$. 

It should be also noted that the arguments presented in \cite{IM} do not apply to the  case where the pressure 
is a constant. Therefore, when $p' = 0$ non-singular stars without heat flow
can be also constrcuted in other versions of the HL theory, although  when $p' \not=0$ this is possible
only in the HMT setup.  In addition, the definitions of the quantities $\rho,\; p$ and $v \; (\equiv q e^{\mu})$ used
in this paper are different from the ones used usually in general relativity. For detail, we refer readers to
Appendix B of \cite{GPW}, specially to Eq.(B16).

\section{Junction Conditions}

\renewcommand{\theequation}{7.\arabic{equation}} \setcounter{equation}{0}

To consider the junction conditions across the hypersurface of a compact object, let us first divide the whole spacetime into three
regions, $V^{\pm}$ and $\Sigma$, where $V^{-} \; (V^{+})$ denotes the internal (external) region of the star, and $\Sigma$ is 
the surface of the star.  As shown in Paper II, once  the metric is cast in the form (\ref{3.1b}), the 
coordinates $t$ and $r$ are all uniquely defined, so that  the coordinates defined in $V^{+}$ and $V^{-}$ must
 be the same,  
$ \big\{x^{+\mu}\big\} = \big\{x^{-\mu}\big\} = (t, r, \theta,\phi)$. 
Since the quadratic terms of the highest derivatives of the metric coefficients $\mu$ and $\nu$ are only terms of the forms,
$\nu''^{2},\; \nu''\nu'''$ and $\mu'^{2}$,  the minimal requirements for these two functions are that   
$\nu(r)$ and $\mu(r)$ are respectively at least $C^{1}$  and $C^{0}$
across the surface $\Sigma$,  
and that  
they are at least $C^{4}$ and $C^{1}$ elsewhere. For detail, we refer readers to \cite{GPW}. Similarly, 
the quadratic terms of the Newtonian pre-potential are only involved with   the forms,
 $\varphi\varphi'',\; \varphi^{2}_{,r}$, and $\varphi\varphi'$. Therefore, the minimal requirment for $\varphi$ is
to be at least $C^{0}$ across the surface $\Sigma$. On the other hand,  the gauge field $A$ 
 and its derivatives all appear linearly. Thus, mathematically it can be even not continuous across  $\Sigma$. However, in this paper
 we shall require that $A$ be  at least $C^{0}$ too across $\Sigma$. Elsewhere,  $A$ and $\varphi$ are at least $C^{1}$. 
Then,  we can write $A$ and $\varphi$ in the form,
\bq
\lb{7.1}
E(r) = E^{+}(r) H\left(r-r_0\right) + E^{-}(r)\left[1 - H\left(r-r_0\right)\right], 
\eq
where $E = (A, \varphi)$, $r_{0}$ is the radius of the star, and $H(x)$ denotes the Heavside function, defined as
\bq
\lb{7.2}
H(x) = \cases{1, & $x > 0$,\cr
0, & $x < 0$.\cr}
\eq
Since $A$ and $\varphi$ are continuous ($C^{0}$) across $r = r_0$, we must have  
\bq
\lb{7.3}
{\mbox{limit}}_{r \rightarrow r_{0}^{+}}{E^{+}(r) } = {\mbox{limit}}_{r \rightarrow r_{0}^{-}}{E^{-}(r)}.
\eq
Then, we find that
\bqn
\lb{7.4}
E'(r)&=& E^{D}_{,r}(r), \nb\\
  E''(r) &=& E^{D}_{,rr}(r) + \left[E'\right]^{-}\delta\left(r -r_0\right), 
\eqn
where  
\bqn
\lb{7.5}
 \left[E'\right]^{-} &\equiv &
 {\mbox{limit}}_{r \rightarrow r_{0}^{+}}{E^{+}_{,r}(r) } -  {\mbox{limit}}_{r \rightarrow r_{0}^{-}}{E^{-}_{,r}(r)},\nb\\
 E^{D}(r)  &\equiv& E^{+}H(r-r_0) + E^{-}\Big[1- H(r-r_0)\Big].
 \eqn
Combining the above with Eq.(6.6) of Paper II, we find that  
\bqn
\lb{7.6a}
{\cal{L}}_{K} &=& {\cal{L}}_{K}^{D},\;\;\; {\cal{L}}_{A} = {\cal{L}}_{A}^{D},\nb\\
{\cal{L}}_{V} &=&  {\cal{L}}_{V}^{ D}  +  {\cal{L}}_{V}^{ Im}\delta(r - r_0), \nb\\
{\cal{L}}_{\varphi} &=&  {\cal{L}}_{\varphi}^{ D}   +  {\cal{L}}_{\varphi}^{Im} \delta(r - r_0),
\eqn
where  
\bqn
\lb{7.6b}
{\cal{L}}^{Im}_{V} &\equiv&   \frac{8g_{7}e^{-6\nu}}{\zeta^{4}r^{3}}   
\Big[2r\nu' - \big(1 - e^{2\nu}\big)\Big]   \left[\nu''\right]^{-},\nb\\
{\cal{L}}_{\varphi}^{Im} &\equiv&  \frac{\varphi e^{-4\nu}}{ r^{2}}   
\Big[\Lambda_{g}r^2 e^{2\nu}   + \big(1 - e^{2\nu}\big)\Big]  \left[\varphi'\right].~~~
\eqn
Setting
\bq
\lb{7.7}
J = J^{D} + J^{Im}\delta(r -r_0),
\eq
where $
J \equiv \{J^{t},\; v, \; J_{\varphi},\; J_{A}\}$,  and
$J^{Im}$ has support only on $\Sigma$,
we find that the Hamiltonian constraint
(\ref{3.3b}) can be written as
 \bqn
 \lb{7.8}
& & \int^{D}{\left( {\cal{L}}_{K} + {\cal{L}}_{V} - {\cal{L}}_{\varphi}^{(1)} - 8 \pi G J^{t} \right) e^{\nu} r^{2} dr}\nb\\
& & ~~~~ =  \frac{1}{4\pi} \Big(8\pi G J^{t, Im} + {\cal{L}}^{Im}_{\varphi} -  {\cal{L}}^{Im}_{V}\Big),
 \eqn
where 
\bq
\lb{7.9}
 \int^{D}{I(r) dr} = {\mbox{limit}}_{\epsilon \rightarrow 0}\left(\int_{0}^{r_0 - \epsilon}{I(r) dr}
 + \int_{r_0 + \epsilon}^{\infty}{I(r) dr}\right).
 \eq
It shold be noted that in writing Eq.(\ref{7.8}), we had used the conversion,
\bq
\lb{7.10}
\int{\sqrt{g}d^{3}x  f(r) \delta(r - r_0)} = f(r_0).
\eq

The momentum constraint (\ref{3.3c}) will take the same form in Regions $V^{\pm}$, while on the surface
$\Sigma$ it yields
\bq
\lb{7.11}
v^{Im} = 0.
\eq
That is, the surface does not support impulsive heat flow  in the radial direction. Similarly,  
Eqs.(\ref{3.3e}) and (\ref{3.3f}) take the same forms in Regions $V^{\pm}$. While on $\Sigma$ they reduce,
respectively, to
\bqn
\lb{7.12a}
& & \Big[\Lambda_{g} r^2 e^{2\nu} + \big(1 - e^{2\nu}\big)\Big]\left[\varphi'\right]^{-} = 8\pi G r^2 e^{4\nu} J_{\varphi}^{Im}, ~~~~~ \\
\lb{7.12b}
& & J_{A}^{Im} = 0,\; (r = r_0).
\eqn

 On the other hand, the dynamical equations (\ref{3.3g}) and (\ref{3.3h}) take the same forms in Regions $V^{\pm}$, 
 and  on the surface $\Sigma$
 they yield, 
 \bqn
 \lb{7.13a}
& &\Bigg\{\varphi e^{-2\nu} \Big[\Lambda_{g} r^2 e^{2\nu} + \big(1 - e^{2\nu}\big)\Big]\left[\varphi'\right]^{-}\nb\\
& & ~~~~~~~~~~~~ + 2 r^{2}F^{\varphi, Im}_{rr}  \Bigg\}\delta(r -r_0) \nb\\
& & ~~ =   -2 r^{2}\Big(F^{Im}_{rr}  + 8\pi G e^{2\nu} p^{Im}_{r}\Big),\\
 \lb{7.13b}
& & \Bigg\{\left[\mu'\right]^{-} + \frac{1}{2} e^{2(\nu-\mu)} {\cal{L}}_{\varphi}^{Im} \nb\\
& & ~~~~  + \frac{e^{2(\nu - \mu)}}{r^{2}}\Big(F^{\varphi, Im}_{\theta\theta} + F^{A, Im}_{\theta\theta}\Big)\Bigg\}\delta(r - r_0)\nb\\
& & ~~ = - \frac{e^{2(\nu - \mu)}}{r^{2}}\Big(F^{Im}_{\theta\theta} + 8\pi Gr^2 p^{Im}_{\theta}\Big),
\eqn
where $F^{Im}_{ij}$ are given by Eq.(A5) in Paper II,  $F^{\varphi, Im}_{ij}$ are given by Eq.(\ref{C.1}) in Appendix C of this paper, and
\bqn
\lb{7.14}
F^{A,Im}_{ij} &=& r^{2}e^{-2\nu} \left[A'\right]^{-}\Omega_{ij},\nb\\
L &=& L^{D} + L^{Im},
\eqn
with $L \equiv (p_{r},\; p_{\theta})$. From Eq.(A5) in Paper II we can see that $L^{Im}$ in general takes the form,
\bq
\lb{7.15}
L^{Im} = L^{(0) Im} \delta(r - r_0) +  L^{(1) Im} \delta'(r - r_0) + L^{(2) Im} \delta''(r - r_0).
\eq
The above represents the general junction conditions of a spherical compact object made of a fluid with heat flow, in which a thin matter  shell 
appears on $\Sigma$. 

In the following we shall consider the matching of the perfect fluid solutions found in Sec. VI to the vacuum ones found in
Sec. V. Since 
\bq
\lb{7.16}
\nu^{\pm} = 0,\;\;\; \varphi^{\pm} = 0,
\eq
we immediately obtain
\bqn
\lb{7.17}
R^{\pm}_{ij} &=& 0,\;\;\;   {\cal{L}}_{V}^{\pm} = 2\Lambda_{\pm},\;\;\;  {\cal{L}}_{V}^{Im} =0,\nb\\
F^{\pm}_{ij} &=& - \Lambda_{\pm}g^{\pm}_{ij}, \;\;\; F^{Im}_{ij} = 0, \nb\\
{\cal{L}}_{\varphi}^{\pm} &=& {\cal{L}}_{\varphi}^{Im} =0, \;\;\;
\left(F_{\varphi}^{\pm}\right)_{ij}   = \left(F_{\varphi}^{Im}\right)_{ij} = 0.  
\eqn
Then, from Eqs.(\ref{7.11}), (\ref{7.12a}), (\ref{7.12b}) and (\ref{7.13a}) we find
\bq
\lb{7.17a}
v^{Im} = J^{Im}_{\varphi} = J^{Im}_{A} = p^{Im}_{r} = 0.
\eq
That is, the radial pressure of the thin shell must vanish. This is similar to what happened in the relativistic case \cite{Santos85}.

In the external region, $V^{+}$, the spacetime is vacuum, 
and the general solutions are given by Eq.(\ref{4.4}),  
\bq
\lb{7.18}
\mu^{+} = \frac{1}{2}\ln\Bigg[\frac{2m}{r} + \frac{1}{3}\Lambda_{+} r^2 - 2A^{+}(r) + \frac{2}{r}\int^{r}{A^{+}(r') dr'}\Bigg],
\eq
for which we have
\bq
\lb{7.19}
{\cal{L}}_{K}^{+}  = \frac{4 A^{+}_{,r}}{r} - 2\Lambda_{+},\;\;\; {\cal{L}}_{A}^{+} = 0.
\eq

In the internal region, $V^{-}$, two classes of solutions of perfect fluid without heat flow are found, given, respectively, by
Eqs.(\ref{5.7a}) and (\ref{5.7b}), which can be written as, 
 \bqn
 \lb{7.20a}
 \mu^{-} &=& \frac{1}{2}\ln\Bigg\{ \frac{1}{3}\Big(\Lambda_{-} - {8\pi G p_{0}}\Big)r^2 - 2A^{-}(r)\nb\\
 & & ~~~~~~~~~  + \frac{2}{r}\int^{r}{A^{-}(r') dr'}\Bigg\},
 \eqn
 for $\Lambda^{-}_{g} = 0$, and
 \bq
 \lb{7.20b} 
 \mu^{-} = \frac{1}{2}\ln\Bigg\{\frac{1}{3}\Big[\Lambda_{-} - \Lambda^{-}_{g}A_{0} - {8\pi G p}\Big]r^2\Bigg\},
 \eq
 for  $A^{-} = A_{0}$, where $p \equiv \Lambda^{-}_{g}A_{0}^{2}/2 + p_{0}$. Then, we find that
 \bqn
 \lb{7.21}
  {\cal{L}}_{K}^{-} &=&\cases{2\big(8\pi G p_{0} - \Lambda_{-}\big) + \frac{4A^{-}_{,r}}{r}, & $ \Lambda_{g}^{-} = 0$,\cr
  2\big(8\pi G p - \Lambda_{-} + \Lambda_{g}^{-}A_{0}\big), & $A^{-} = A_{0}$,\cr}\nb\\
  {\cal{L}}_{A}^{-} &=& 2\Lambda_{g}^{-} A^{-}(r).
    \eqn
 To further study the junction conditions, let us consider the two cases $\Lambda_{g}^{-} = 0$ and
 $A^{-} = A_{0}$ separately.
 
 \subsection{$\Lambda_{g}^{-} = 0$}
 
 In this case, the continuity  conditions of $\mu$ and $A$ across $\Sigma$ read,
 \bqn
 \lb{7.22}
& &  {m} + \frac{1}{6}\Delta\Lambda r^{3}_{0} + \int^{r_{0}}{\Delta{A}(r) dr} = - \frac{4\pi G }{3} p_{0} r^{3}_{0},\nb\\
& &  A^{+}(r_{0}) = A^{-}(r_{0}),
 \eqn
where $ \Delta\Lambda \equiv \Lambda_{+} - \Lambda_{-}$ and $\Delta{A} = A^{+} - A^{-}$.
Then, the Hamiltonian constraint (\ref{7.8}) becomes,
\bq
\lb{7.23}
\int^{r_{0}}_{0}{\tilde{\rho}(r) dr} + \int^{\infty}_{r_{0}}{r A^{+}_{,r}(r) dr} = \frac{1}{2} G J^{t, Im},
\eq
while the dynamical equation (\ref{7.13b}) reduces to,
\bq
\lb{7.24}
\Delta\Lambda = - 8 \pi G \Big(p_{0} + 2 p^{(0) Im}_{\theta}\Big),
\eq
where $p^{Im}_{\theta} \equiv p^{(0) Im}_{\theta}\delta(r-r_{0})$ [cf. Eq.(\ref{7.15})]. 

When the matter thin shell
does not exist, we must set $J^{t, Im} = p^{(0) Im}_{\theta} = 0$, and Eqs.(\ref{7.22})-(\ref{7.24}) become the matching conditions 
for the constants $\Lambda_{\pm},\; m,\; p_{0}$ and the functions $A^{\pm}(r)$ and $\tilde{\rho}(r)$.

 \subsection{$A = A_{0}$}
 
 In this case, it can be shown that the continuity  conditions for $\mu$ and $A$  become,
 \bqn
 \lb{7.22a}
& &  {m} + \frac{1}{6}\Delta\Lambda r^{3}_{0} + \int^{r_{0}}{{A}^{+}(r) dr} = r_{0}A_{0} \nb\\
&& ~~~~~~~~~~~~~~~~~~~~ - \frac{1}{6}\big(\Lambda^{-}_{g}A_{0} + 8\pi G  p\big) r^{3}_{0},\nb\\
& &  A^{+}(r_{0}) = A_{0},
 \eqn
while the Hamiltonian constraint (\ref{7.8}) reduces to,
\bq
\lb{7.23a}
\int^{r_{0}}_{0}{\tilde{\rho}(r) dr} + 4 \int^{\infty}_{r_{0}}{r A^{+}_{,r}(r) dr} = {2} G J^{t, Im}.
\eq
The dynamical equation (\ref{7.13b}), on the other hand,  yields,
\bq
\lb{7.24a}
\Delta\Lambda =  - \Lambda^{-}_{g}A_{0} - 8 \pi G \Big(p + 2 p^{(0) Im}_{\theta}\Big). 
\eq
 
In all the above cases, one can see that the matching is possible even without a thin matter
shell on the surface of the star, $J^{t, Im} = 0 = p^{(0) Im}_{\theta}$, by properly choosing the
free parameters.

 \section{Conclusions}
 
\renewcommand{\theequation}{8.\arabic{equation}} \setcounter{equation}{0}

 In this paper, we have systematically studied spherically symmetric static spacetimes generally
filled with an anisotropic 
fluid with heat flow along the radial direction. When the spacetimes are vacuum, we have found  solutions,
given explicitly by Eqs.(\ref{4.1}) and (\ref{4.4}), from which one can see that the solution is not unique, because the gauge 
field $A$ is  undetermined. When $A = 0$, the solutions reduce to the Schwarzschild (anti-) de Sitter solution.
We have also studied the solar system tests, and found the constraint on the choice of  $A$.
 
 It should be noted that we have adopted a different point of view  of the gauge 
 field $A$ in the IR limit than that adopted in \cite{HMT}. In this paper we have considered it as independent from the
4D metric $g_{\mu\nu}$, although it interacts with them through the field equations. This is 
 quite similar to the Brans-Dicke (BD) scalar field in the BD theory, where the scalar field represents a degree of freedom
 of gravity,  is independent of the metric, and its effects to the spacetime are only through the field equations  \cite{BD}. 
 On the contrary, in \cite{HMT} the authors considered the gauge field $A$ as a part of the lapse function,
 $g_{tt} \simeq - (N - A)^2$ in the IR limit.

 We have also investigated anisotropic fluids with heat flow, and found   perfect fluid solutions, given 
 by Eq.(\ref{5.7a}). By properly choosing the gauge field $A$, the solutions can be free of
spacetime singularities at the center. This is in contrast to other versions of the HL theory \cite{IM}, due to 
the coupling of the fluid with the gauge field. We then have considered two particular cases, in 
which  the pressure is a constant,  quite similar to the Schwarzschild perfect fluid solution. In all these cases, the
spacetimes are free of singularities at the center. 
 
 For a compact object, the spacetime outside of it  is vacuum, matching conditions
 are needed across the surface of the star. With the minimal requirement that the junctions be mathematically meaningful,
 we have worked out the general matching conditions, given by Eqs.(\ref{7.8}) and (\ref{7.11})-(\ref{7.13b}), 
  in which a thin matter shell in general appears on the surface of the star. Applying them to the perfect
  fluids, where the spacetime outside is described by the vacuum solutions (\ref{4.4}), we have found
  the matching conditions in terms of the free parameters of the solutions. When the thin shell is removed, these 
  conditions can also be satisfied by properly choosing the free parameters.
  
    Finally, we note that da Silva argued, in the HMT setup, that the coupling constant $\lambda$ can still be different from one
  \cite{Silva}. If this is indeed the case,  then one might be concerned with  the   strong coupling problem found in other versions of the
  HL theory \cite{BPS,WWb,SC}. However, since the spin-0 graviton is eliminated completely here, as shown explicitly in \cite{HMT,WW,Silva},
   this question is automatically solved in the HMT setup even with $\lambda \not= 1$ \cite{Thanks}. It should be noted that 
   da Silva considered only perturbations of the case with detailed balance condition, and found that the spin-0 mode is not propagating. 
   It is not clear if it is also true for the case without detailed balance. The problem certainly deserves further investigations.

  {\bf Note Added:} A preprint \cite{AP} appeared in arXiv almost simultaneously with ours. These
  authors also studied spherically symmetric static vacuum solutions similar to those in Sec. IV, but  
  did not consider the subjects presented in the other parts of this paper. 
  
~\\{\bf Acknowledgements:}  One of the authors (AW) would like to thank Shinji Mukohyama, Tony Padilla, and Thomas P. Sotiriou for valuable
comments and suggestions. The work of AW was supported in part by DOE  Grant, DE-FG02-10ER41692.

\section*{Appendix A:  Functions $\left(F_{s}\right)_{ij}$ and $F_{(\varphi, n)}^{ij}$ }
\renewcommand{\theequation}{A.\arabic{equation}} \setcounter{equation}{0}

The geometric 3-tensors  $F^{ij}$ and $F_{(\varphi, n)}^{ij}$ defined in  Eq.(\ref{eq3a})
are given  by
  \bqn \lb{A.1}
\left(F_{0}\right)_{ij} &=& - \frac{1}{2}g_{ij},\nb\\
\left(F_{1}\right)_{ij} &=& R_{ij}- \frac{1}{2}Rg_{ij},\nb\\
\left(F_{2}\right)_{ij} &=& 2\left(R_{ij} -
\nabla_{i}\nabla_{j}\right)R
-  \frac{1}{2}g_{ij} \left(R - 4\nabla^{2}\right)R,\nb\\
\left(F_{3}\right)_{ij} &=& \nabla^{2}R_{ij} - \left(\nabla_{i}
\nabla_{j} - 3R_{ij}\right)R - 4\left(R^{2}\right)_{ij}\nb\\
& & +  \frac{1}{2}g_{ij}\left( 3 R_{kl}R^{kl} + \nabla^{2}R
- 2R^{2}\right),\nb\\
\left(F_{4}\right)_{ij} &=& 3 \left(R_{ij} -
\nabla_{i}\nabla_{j}\right)R^{2}
 -  \frac{1}{2}g_{ij}\left(R  - 6 \nabla^{2}\right)R^{2},\nb\\
 \left(F_{5}\right)_{ij} &=&  \left(R_{ij} + \nabla_{i}\nabla_{j}
 \right) \left(R_{kl}R^{kl}\right)
 + 2R\left(R^{2}\right)_{ij} \nb\\
& & + \nabla^{2}\left(RR_{ij}\right) - \nabla^{k}\left[\nabla_{i}
\left(RR_{jk}\right) +\nabla_{j}\left(RR_{ik}\right)\right]\nb\\
& &  -  \frac{1}{2}g_{ij}\left[\left(R - 2 \nabla^{2}\right)
\left(R_{kl}R^{kl}\right)\right.\nb\\
& & \left.
- 2\nabla_{k}\nabla_{l}\left(RR^{kl}\right)\right],\nb\\
\left(F_{6}\right)_{ij} &=&  3\left(R^{3}\right)_{ij}  +
\frac{3}{2}
\left[\nabla^{2}\left(R^{2}\right)_{ij} \right.\nb\\
 & & \left.
 - \nabla^{k}\left(\nabla_{i}\left(R^{2}\right)_{jk} + \nabla_{j}
 \left(R^{2}\right)_{ik}\right)\right]\nb\\
 & &    -  \frac{1}{2}g_{ij}\left[R^{k}_{l}R^{l}_{m}R^{m}_{k} -
 3\nabla_{k}\nabla_{l}\left(R^{2}\right)^{kl}\right],\nb\\
 \left(F_{7}\right)_{ij} &=&  2 \nabla_{i}\nabla_{j}
 \left(\nabla^{2}R\right) - 2\left(\nabla^{2}R\right)R_{ij}\nb\\
 & &    + \left(\nabla_{i}R\right)\left(\nabla_{j}R\right)
  -  \frac{1}{2}g_{ij}\left[\left(\nabla{R}\right)^{2} +
  4 \nabla^{4}R\right],\nb\\
\left(F_{8}\right)_{ij} &=&  \nabla^{4}R_{ij} -
\nabla_{k}\left(\nabla_{i}\nabla^{2} R^{k}_{j}
                            + \nabla_{j}\nabla^{2} R^{k}_{i}
                            \right)\nb\\
& & - \left(\nabla_{i}R^{k}_{l}\right)
\left(\nabla_{j}R^{l}_{k}\right)
       - 2 \left(\nabla^{k}R^{l}_{i}\right) \left(\nabla_{k}R_{jl}
       \right)\nb\\
& &    -  \frac{1}{2}g_{ij}\left[\left(\nabla_{k}R_{lm}\right)^{2}
        -
        2\left(\nabla_{k}\nabla_{l}\nabla^{2}R^{kl}\right)\right],\\
  \lb{A.2}
F_{(\varphi, 1)}^{ij} &=& \frac{1}{2}\varphi\left\{\Big(2K + \nabla^{2}\varphi\Big) R^{ij}  
- 2 \Big(2K^{j}_{k} + \nabla^{j} \nabla_{k}\varphi\Big) R^{ik} \right.\nb\\
& & ~~~~~ - 2 \Big(2K^{i}_{k} + \nabla^{i} \nabla_{k}\varphi\Big) R^{jk}\nb\\
& &~~~~~\left. 
- \Big(2\Lambda_{g} - R\Big) \Big(2K^{ij} + \nabla^{i} \nabla^{j}\varphi\Big)\right\},\nb\\
F_{(\varphi, 2)}^{ij} &=& \frac{1}{2}\nabla_{k}\left\{\varphi{\cal{G}}^{ik}  
\Big(\frac{2N^{j}}{N} + \nabla^{j}\varphi\Big) \right. \nb\\
& & \left.
+ \varphi{\cal{G}}^{jk}  \Big(\frac{2N^{i}}{N} + \nabla^{i}\varphi\Big) 
-  \varphi{\cal{G}}^{ij}  \Big(\frac{2N^{k}}{N} + \nabla^{k}\varphi\Big)\right\}, \nb\\   
F_{(\varphi, 3)}^{ij} &=& \frac{1}{2}\left\{2\nabla_{k} \nabla^{(i}f^{j) k}_{\varphi}  
- \nabla^{2}f_{\varphi}^{ij}   - \left(\nabla_{k}\nabla_{l}f^{kl}_{\varphi}\right)g^{ij}\right\},\nb\\
\eqn
where
\bqn
\lb{A.3}
f_{\varphi}^{ij} &=& \varphi\left\{\Big(2K^{ij} + \nabla^{i}\nabla^{j}\varphi\Big) 
- \frac{1}{2} \Big(2K + \nabla^{2}\varphi\Big)g^{ij}\right\}.\nb\\
\eqn

\section*{Appendix B:  Function   $F_{(\varphi, n)}^{ij}$ in Spherical Static Spacetimes}
\renewcommand{\theequation}{B.\arabic{equation}} \setcounter{equation}{0}

In the spherically symmetric static  spacetimes described by the metric (\ref{3.1b}), the function  
 $F_{(\varphi, n)}^{ij}$ defined by Eq.(\ref{A.2}) is given  by
\bqn
\lb{B.2}
\left[F_{(\varphi, 1)}\right]_{ij} &=&\frac{\varphi e^{-2\nu}}{r^{2}}\Bigg\{\Big[e^{2\nu}\Big(1-\Lambda_{g}r^{2}\Big) - \big(1 + r\nu'\big)\Big]\varphi''\nb\\
& & ~~ + \Big[e^{2\nu}\Big(\Lambda_{g}r^{2} -1\Big)  + \big(3 + r\nu'\big)\Big]\nu' \varphi'\nb\\
& &~~  - 2e^{\mu + \nu}\Big[e^{2\nu}\Big(\Lambda_{g}r^{2} -1\Big)+ 1\Big]\mu'\nb\\
& &~~  + 2e^{\mu+\nu}\left(2 - r\mu'\right) \nu' \Bigg\}\delta^{r}_{i}\delta^{r}_{j}\nb\\
& &  + \frac{1}{2}\varphi e^{-4\nu}\Bigg\{\Big[r\nu' - \Big(1- e^{2\nu}\Big)\Big]\varphi'' \nb\\
& &~~ + \Big(3 - r\nu' - e^{2\nu}\Big)\nu' \varphi' - 2 \Lambda_{g}r e^{2\nu} \varphi'\nb\\
& & ~~ - 2e^{\mu+\nu}\Big(2 - r\nu' - e^{2\nu}\Big)\mu' \nb\\
& &~~ + 4e^{\mu+\nu}\Big(\nu' - \Lambda_{g}r e^{2\nu}\Big)\Bigg\}\Omega_{ij},\nb\\
\left[F_{(\varphi, 2)}\right]_{ij} &=&\frac{ e^{-2\nu}}{2r^{2}}\Bigg\{\Big[e^{2\nu}\Big(\Lambda_{g}r^{2}-1\Big) + 1\Big]\Big[\varphi\varphi''
\nb\\
& & ~~  + \Big(\varphi' + \varphi\nu'\Big)\varphi' + 2 e^{\mu+\nu} \Big(\varphi' + \varphi\mu'\Big)\Big] \nb\\
& &~~ + 4\varphi e^{\mu+\nu}\Big(\nu' - \Lambda_{g}r e^{2\nu}\Big)\nb\\
& &~~  - 2\Lambda_{g}r\varphi e^{2\nu}\varphi'\Bigg\} \delta^{r}_{i}\delta^{r}_{j}\nb\\
& &  + \frac{1}{2} e^{-4\nu}\Bigg\{r\varphi\Big(\nu' - \Lambda_{g}r e^{2\nu}\Big)\varphi'' \nb\\
& &~~ + r\varphi\Big(\varphi' + 2 e^{\mu +\nu}\Big)\nu''\nb\\
&& ~~ +   r\Big(\nu' - \Lambda_{g}r e^{2\nu}\Big)\varphi^{'2} \nb\\
& & ~~ - r\varphi\Big(3\varphi' + 4 e^{\mu +\nu}\Big)\nu^{'2}\nb\\
& & ~~ -  \Big(\varphi - 2r e^{\mu + \nu}  - \Lambda_{g} r^{2}\varphi e^{2\nu}\Big)\nu' \varphi'\nb\\
& &~~ - 2e^{\mu+\nu}\Big(\Lambda_{g}r^{2} e^{2\nu}\varphi' + \varphi\nu'\Big)\nb\\
& & ~~ + 2r\varphi e^{\mu + \nu}\Big(\nu' - \Lambda_{g}r e^{2\nu}\Big)\mu'\Bigg\}\Omega_{ij},\nb\\
\left[F_{(\varphi, 3)}\right]_{ij} &=&\frac{ e^{-2\nu}}{r^{2}}\Bigg[r\varphi\nu'\varphi'' + \varphi^{'2}
- \varphi\big(r\nu' + 2\big)\nu'\varphi'\nb\\
& & ~~ + 2e^{\mu +\nu}\Big(\varphi' + r\varphi\nu'\mu' -2\varphi\nu'\Big)\Bigg] \delta^{r}_{i}\delta^{r}_{j}\nb\\
& &  + \frac{1}{2} e^{-4\nu}\Bigg\{2r\Big(\varphi' - \varphi\nu' + e^{\mu + \nu}\Big)\varphi'' \nb\\
& &~~ -  r\varphi\Big(\varphi' + 2 e^{\mu +\nu}\Big)\nu'' \nb\\
& & ~~ + \Big[4r\big(\varphi\nu' -\varphi'\big) + \varphi\Big]\nu'\varphi'\nb\\
&& ~~ + e^{\mu +\nu}\Big[4r\varphi\big(\nu' - \mu'\big)\nu' + 2\varphi\nu' \nb\\
&& ~~~~~~~~~~~~~ + 2r\big(\mu' -3\nu'\big)\varphi'\Big]\Bigg\}\Omega_{ij}.
\eqn

\section*{Appendix C:  Impulsive Parts of     $F_{(\varphi, n)}^{ij}$}
\renewcommand{\theequation}{C.\arabic{equation}} \setcounter{equation}{0}

From Eqs.(\ref{7.1}) and (\ref{7.4}) we find that $\left[F_{(\varphi, n)}\right]_{ij}$ given by
Eq.(\ref{B.2}) takes the form,
\bq
\lb{C.0}
\left[F_{(\varphi, n)}\right]_{ij} = \left[F_{(\varphi, n)}\right]_{ij}^{D} + \left[F^{Im}_{(\varphi, n)}\right]_{ij}\delta(r-r_0),
\eq
where
\bqn
\lb{C.1}
\left[F^{Im}_{(\varphi, 1)}\right]_{ij} &=&\frac{\varphi e^{-2\nu}}{r^{2}}\Bigg[e^{2\nu}\Big(1-\Lambda_{g}r^{2}\Big) - \big(1 + r\nu'\big)\Bigg] \nb\\
& & ~~~~~~~~~\times
\left[\varphi'\right]^{-}\delta^{r}_{i}\delta^{r}_{j}\nb\\
& &  + \frac{1}{2}\varphi e^{-4\nu}\Bigg[r\nu' - \Big(1- e^{2\nu}\Big)\Bigg]\left[\varphi'\right]^{-}\Omega_{ij},\nb\\
\left[F^{Im}_{(\varphi, 2)}\right]_{ij} &=&\frac{\varphi e^{-2\nu}}{2r^{2}}\Bigg[e^{2\nu}\Big(\Lambda_{g}r^{2}-1\Big) + 1\Bigg]
\left[\varphi'\right]^{-} \delta^{r}_{i}\delta^{r}_{j}\nb\\
& &  + \frac{1}{2}r\varphi  e^{-4\nu}\Bigg[\Big(\nu' - \Lambda_{g}r e^{2\nu}\Big)\left[\varphi'\right]^{-} \nb\\
& & ~~~~~~~~
  + \Big(\varphi' + 2 e^{\mu +\nu}\Big)\left[\nu'\right]^{-}\Bigg]\Omega_{ij},\nb\\
\left[F^{Im}_{(\varphi, 3)}\right]_{ij} &=&\frac{\varphi\nu'  e^{-2\nu}}{r} \left[\varphi'\right]^{-} \delta^{r}_{i}\delta^{r}_{j}\nb\\
& &  + \frac{1}{2} r e^{-4\nu}\Bigg[2\Big(\varphi' - \varphi\nu' + e^{\mu + \nu}\Big)\left[\varphi'\right]^{-} \nb\\
& &~~ -  \varphi\Big(\varphi' + 2 e^{\mu +\nu}\Big)\left[\nu'\right]^{-} \Bigg]\Omega_{ij}.
\eqn

\end{document}